\documentclass{IEEEtran}
\usepackage{cite}
\usepackage[utf8]{inputenc}
\usepackage{amsmath}
\usepackage{amssymb}
\usepackage{graphicx}
\begin{document}

\title{Plane-wave compounding with adaptive joint coherence factor weighting}
\author{Nikunj~Khetan, Jerome Mertz
\thanks{This work was partially supported by the National Institutes of Health (R21GM134216) and by the Boston University Photonics Center.}
\thanks{N. K. is with Boston University Mechanical Engineering, 110 Cummington Mall, Boston, MA 02215 (e-mail: nkhetan@bu.edu). }
\thanks{J. M. is with Boston University Biomedical Engineering, 44 Cummington Mall, Boston, MA 02215 (e-mail: jmertz@bu.edu).}}

\maketitle

\begin{abstract}
Coherent Plane Wave Compounding (CPWC) is widely used for ultrasound imaging. This technique involves sending plane waves into a sample at different transmit angles and recording the resultant backscattered echo at different receive positions. The time-delayed signals from the different combinations of transmit angles and receive positions are then coherently summed to produce a beamformed image. Various techniques have been developed to characterize the quality of CPWC beamforming based on the measured coherence across the transmit or receive apertures. Here, we propose a more fine-grained approach where the signals from every transmit/receive combination are separately evaluated using a quality metric based on their joint spatio-angular coherence. The signals are then individually weighted according to their measured Joint Coherence Factor (JCF) prior to being coherently summed. To facilitate the comparison of JCF beamforming compared to alternative techniques, we further propose a method of image display standardization based on contrast matching. We show results from tissue-mimicking phantoms and human soft-tissue imaging. Fine-grained JCF weighting is found to improve CPWC image quality compared to alternative approaches. 
\end{abstract}            

\section{Introduction}

Ultrasound imaging is widely used in modern healthcare because it is non-invasive, low cost and broadly applicable. In most cases, ultrasound image reconstruction is based on the line-by-line application of a simple delay-and-sum (DAS) algorithm to the received echo signals, limiting imaging speeds to roughly video rate. However, the introduction of coherent plane wave compounding (CPWC) \cite{montaldo_coherent_2009} has led to a huge advance in the field by enabling unprecedented speeds with little compromise in image quality. Already, CPWC is used in a variety of clinical applications, such as shear wave imaging \cite{montaldo_coherent_2009} and Doppler-based flow imaging \cite{bercoff_ultrafast_2011}.

However, CPWC is not without its limitations. Because it is a multiplexing technique, it is generally more susceptible to noise than line-by-line techniques making use of focused transmit beams. Several sources of noise can degrade signal in CPWC, such as off-axis or out-of-focus clutter, deviations in the local speed of sound caused by sample inhomogeneities, multiple scattering within the sample or from subresolution features resulting in speckle noise, uncorrelated electronic noise, etc.. In addition, because CPWC requires data obtained from several successive plane-wave insonifications, any inter-frame micro-motion within the sample, random or otherwise, can induce error in the CPWC reconstruction.

A key advance in the implementation of CPWC has been the recognition that not all signals are of high quality. This has led to the formulation of a variety of metrics to characterize signal quality, which can then be used to adaptively optimize the reconstruction of images based on the incoming signals themselves. A common metric is the coherence factor \cite{long_spatial_2022}, which measures the self-similarity of signals across a particular dimension. A high degree of self-similarity, or correlation, is indicative of high-quality signal whereas a low degree is indicative of noise or clutter, which is to be suppressed.

Early approaches in ultrasound imaging adopted the concept of a spatial coherence factor (CF) \cite{mallart_van_1991}, wherein self-similarity is measured along the dimension of the receive aperture space. In this case, the CF is defined as the ratio of coherent to incoherent signal power across the aperture transducer elements. While initially the spatial CF was used simply to characterize image quality, it was quickly realized that it could be applied as a filter to adaptively suppress uncorrelated noise or clutter \cite{mallart_adaptive_1994}. Successive variations of the CF came to being. For example, the generalized CF (GCF) is a metric based on the spatial frequency content of the signal, given by the ratio of the signal power below a cutoff frequency to the total power \cite{li_adaptive_2003}. Alternatively, the phase coherence factor (PCF) is a measure of the phase variations across the aperture \cite{camacho_phase_2009}.

While the spatial CF is essentially a measure of the normalized variance of signal strengths across the receive aperture \cite{hollman_coherence_1999}, more comprehensive information is provided by signal covariance. For example, the short lag spatial coherence (SLSC) \cite{lediju_short-lag_2011,lediju_bell_short-lag_2013} provides a measure of the signal covariance strength up to a cutoff lag, where a short cutoff lag plays a similar role as a high cutoff frequency in GCF. The delay multiply and sum (DMAS) algorithm \cite{matrone_delay_2015}, provides a measure of covariance strength over all lags, making it similar to SLSC (though normalized differently and without an added step of local time averaging). Different from CF approaches, SLSC and DMAS generally provide final images rather than serving as filter-functions that are subsequently applied to DAS-based image reconstruction.

More recently, it has been recognized that coherence in the receive aperture can equally be applied to coherence in the transmit aperture, owing to the principle of acoustic reciprocity \cite{li_angular_2017}. When applied to CPWC, this led to the concept of short lag angular coherence (SLAC) \cite{li_angular_2017}, which is the analog of SLSC but with covariance measured along the transmit coordinate (angle) rather than the receive coordinate (space). It has also led to the concept of an angular CF, and ultimately to the concept of a united CF (UCF) \cite{yang_coherent_2020} that aims to combine signal variance information along both angle and space coordinates. However, as will be shown below, UCF does not fully exploit this information.

Of course, other filtering strategies are possible. Notably, the Capon minimum variance (MinVar) algorithm derives the filter weights that optimally suppress clutter noise while preserving signal gain \cite{synnevag_benefits_2009}. Indeed, efforts have been made to apply MinVar weighting across both spatial and angular coordinates  \cite{rindal_double_2016,zhao_plane_2015,nguyen_spatial_2018}, though with a drawback that MinVar weights are derived according to expectations of signal covariances. These expectations must be estimated from the actual measured signal covariances, often requiring signal pooling that can undermine the benefits of MinVar weighting in the first place. In general, the MinVar algorithm must be applied with care, since small errors in the expected covariances can easily lead to worse outcomes in image quality.

In this work, we re-examine the application of CF-based filtering to CPWC. We propose a new filtering algorithm that is inspired by previous CF methods but goes a step further in how fine-grained the level of filtering is taken. Rather than consider a spatial or angular CF separately, or try to amalgamate these into a single weighting factor \cite{yang_coherent_2020}, we consider every possible combination of transmit to receive sound trajectories. Our premise is that different trajectories yield different signal qualities depending on the different uncontrollable influences they are subjected to (clutter, inhomogeneities, scattering, noise, etc.). That is, to each transmit/receive trajectory we assign a quality metric specific to that trajectory, based on the measured signals themselves. By suppressing the low quality signals at the fine-grained level of each individual trajectory, we thus suppress their deleterious contributions.

This paper is organized as follows. In section 2, we describe how the individual trajectory weights are calculated based on the measured signals, using an adaptive joint coherence factor (JCF) algorithm applied to CPWC. We also introduce a method of contrast matching to standardize the comparison between images. In section 3, we compare the results of JCF weighting with other popular weighting strategies, using datasets collected from both phantom and human imaging. As will be shown, JCF weighting compares favorably to other strategies.

\section{Theory and Methods}

We consider here a linear transducer array of $N$ elements with position $x_{n}$ along the azimuthal direction. CPWC is performed by transmitting a set of $M$ plane waves into the sample, each with different steering angle $\theta_{m}$. The resulting echo signals associated with any given target location $(x,z)$ within the sample, appropriately time delayed, are thus obtained in the form of a $M\times N$ matrix, with element $s_{m,n}$ corresponding to the signal obtained from a plane wave transmit of steering angle $\theta_{m}$ and received on array element $x_{n}$ :
\begin{equation}
\mathbf{S}\left(  x,z\right)  =\left[
\begin{array}
[c]{cccc}%
s_{1,1} & s_{1,2} & ... & s_{1,N}\\
s_{2,1} & s_{2,2} & ... & s_{2,N}\\
... & ... & ... & ...\\
s_{M,1} & s_{M,2} & ... & s_{M,N}%
\end{array}
\right]
\end{equation}

That is, $\mathbf{S}(x,z)$ encompasses the signal elements obtained from every combination of transmit angles and receive positions, corresponding to different sound trajectories. We assume these signals are complex (i.e. they are the complex analytical representations of the measured real signals), and henceforth omit the dependence of $\mathbf{S}$ on $(x,z)$, taking this to be implicit.

\subsection{Beamforming with JCF weighting}

In the conventional DAS implementation of CPWC, the beamformed field associated with a sample target location is given by the simple summation of the signal elements in $\mathbf{S}$ (generally weighted by depth-dependent apodization factors, which we omit here for brevity). That is,
\begin{equation}
B_{DAS}\left(  x,z\right)  =\frac{1}{MN}\sum\limits_{m=1}^{M}\sum
\limits_{n=1}^{N}s_{m,n}%
\end{equation}

Our basic strategy is not to treat each signal element $s_{m,n}$ equally, as in standard DAS beamforming, but rather to adaptively assign a weighting factor $w_{m,n}$ to each individual signal element based on its measured quality. That is, our beamforming is given by
\begin{equation}
B_{JCF}\left(  x,z\right)  =\frac{1}{MN}\sum\limits_{m=1}^{M}\sum
\limits_{n=1}^{N}w_{m,n}s_{m,n}%   
\end{equation}

The metric we use here to define signal quality is based on a joint spatio-angular coherence factor (JCF for short), defined by 
\begin{equation}
\label{JCF_weight_basic}
w_{m,n}\left(  x,z\right)  =\frac{\left\vert \sum\limits_{m^{\prime}=1}%
^{M}\sum\limits_{n^{\prime}=1}^{N}s_{m^{\prime},n}s_{m,n^{\prime}}\right\vert
^{\alpha}}{\left(  MN\right)  ^{\alpha-1}\sum\limits_{m^{\prime}=1}^{M}%
\sum\limits_{n^{\prime}=1}^{N}\left\vert s_{m^{\prime},n}\right\vert ^{\alpha
}\left\vert s_{m,n^{\prime}}\right\vert ^{\alpha}}%   
\end{equation}

\noindent where $\alpha\geq0$ is a user-defined ``smoothness" parameter whose role will be made clear later (note that JCF beamforming reduces to conventional DAS beamforming when $\alpha=0$). Equation \eqref{JCF_weight_basic} has a simple interpretation: $w_{m,n}$ corresponds to the coherence of $s_{m,n}$ measured jointly along both the spatial (associated row) and angular (associated column) directions. $w_{m,n}$ ranges between $0$ and $1$, indicating low and high degrees of coherence respectively. Signal elements with low coherence are considered corrupted and of low quality, and suppressed accordingly. We emphasize here that $w_{m,n}$ is a two  dimensional matrix, as opposed to the weighting factors used in most filtering-based techniques which are one dimensional vectors, or scalar in UCF (see Tables \ref{tableEQ} and \ref{tableEQ_part2}). In other words, $w_{m,n}$ weights each signal element individually, as opposed to weighting entire rows or columns of signal elements, or the entire matrix of signal elements. As will be shown below, this leads to considerable improvements in image quality.

\begin{figure*}[!t]
\centering
\includegraphics[width= \textwidth]{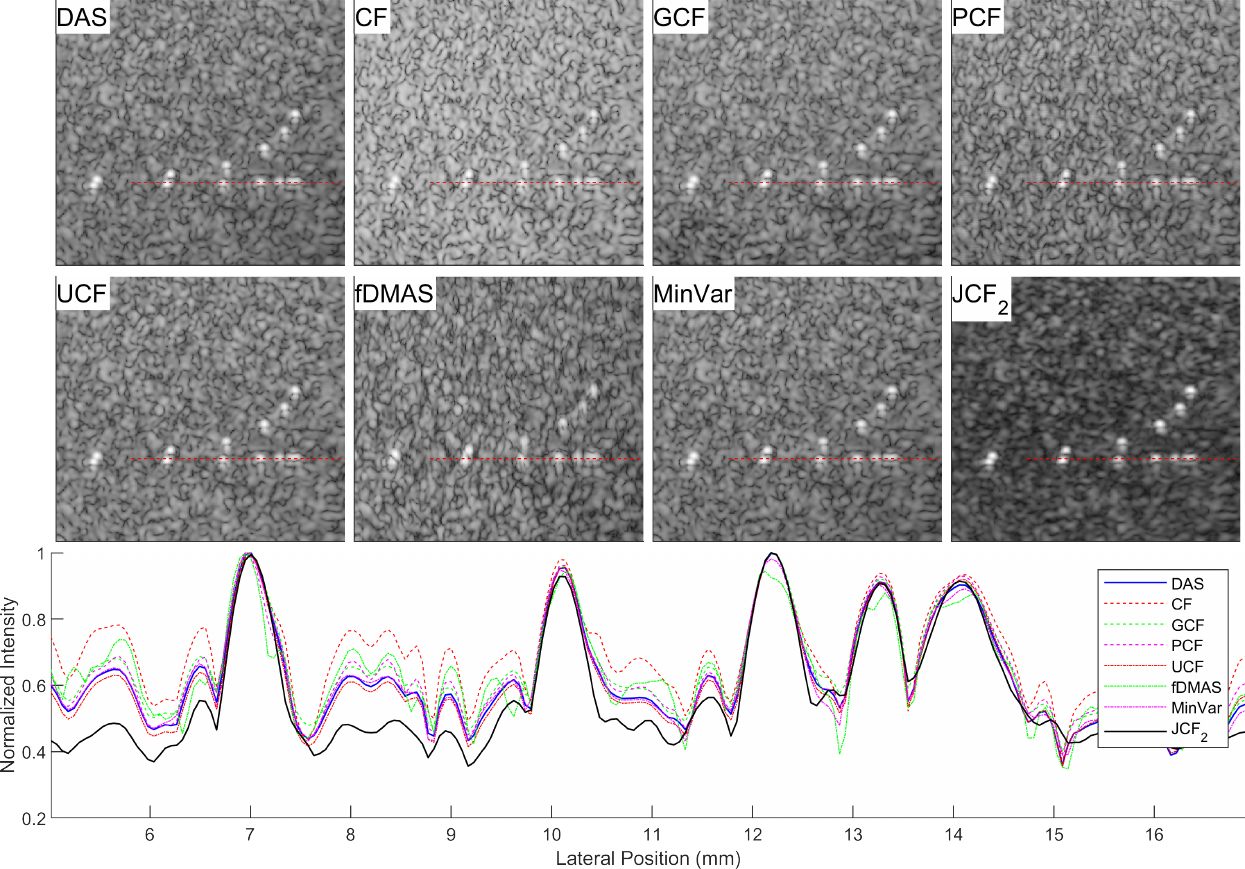}
\caption{Top: Contrast-matched images of resolution targets in a CIRS GSE 040 phantom obtained at ~30 mm depth. Beamforming reconstruction method is indicated in panels, where JCF$_2$ indicates $\alpha=2$. Image size is 16 mm x 15 mm (Horizontal x Vertical). Bottom: Intensity plots across red lines in panels. JCF suppresses speckle noise without noticeably affecting resolution.}
\label{figRes}
\end{figure*}

\subsection{Contrast matching}

To properly assess the benefits of JCF beamforming compared to alternative methods, we must first standardize what we mean by image quality. Common definitions of image quality involve the association of a contrast ratio (CR) \cite{smith_frequency_1985} or a contrast-to-noise ratio (CNR) \cite{patterson_improvement_1983} to the image. These definitions compare signal with background (or background noise), meaning they require an a priori delineation between what is understood to be signal and what is understood to be background. When imaging phantom samples of known inclusions, such a delineation is clear; however, when imaging samples with unknown inclusions where the distinction between signal and background is not so clear cut this delineation becomes decidedly less straightforward. Exacerbating this difficulty is the problem that ultrasound signals can span an enormously large dynamic range, meaning that some form of intensity compression is required for the display of images. The simplest and most common approaches involve log or gamma compression, though more sophisticated approaches can also be applied \cite{hverven_point_2017,rindal_effect_2019}. However, it is well known that both CR and CNR vary considerably with different compression algorithms, meaning they are not reliable indicators of image quality to start with. This has led to the formulation of a generalized CNR (gCNR) \cite{rodriguez-molares_generalized_2020} which characterizes the separability between signal and background in a compression-agnostic manner. However, the gCNR metric requires yet additional a priori information, not only about the binary distinction between signal and background (usually requiring manual intervention), but also about the respective intensity statistics of each. Such information, again, is difficult if not impossible to obtain when imaging real samples. In other words, a quantitative metric for ultrasound image quality has proven surprisingly difficult to define \cite{rindal_very_2023}.

In this work, we fully acknowledge this difficulty. Nevertheless, our aim is to provide side-by-side comparisons of images obtained with different beamforming methods. To ensure that these comparisons are made with images on equal footing, we introduce here a strategy of contrast matching wherein the overall contrasts of all images are adjusted to be the same upon display. In particular, we choose the contrast of the DAS image to serve as the reference contrast. In our case, the DAS image is displayed with commonly used gamma compression, using a gamma factor set here to $\gamma_{DAS}=0.25$. That is, throughout this work DAS images are displayed with pixel values given by
\begin{equation}
I_{DAS}\left(  x,z\right)  =\left\vert B_{DAS}\left(  x,z\right)  \right\vert
^{\gamma_{DAS}=0.25}%    
\end{equation}

We define the contrast of this DAS image to be \cite{goodman_speckle_2007}
\begin{equation}
\label{contrast}
K_{DAS}=\frac{\sigma_{DAS}}{\left\langle I_{DAS}\right\rangle \,}%
\end{equation}
\noindent where $\left\langle I_{DAS}\right\rangle $ and $\sigma_{DAS}$ correspond respectively to the mean and standard deviation of the pixel values across the displayed image. That is, $K_{DAS}$ is a measure of the relative range of pixel variations throughout the displayed DAS image. This measure has the advantage that it makes no attempt to categorize pixels as belonging to signal or background, and thus requires no a priori information or manual intervention. As an example, let us consider an image obtained with JCF beamforming. The displayed JCF image is given by
\begin{equation}
I_{JCF}\left(  x,z\right)  =\left\vert B_{JCF}\left(  x,z\right)  \right\vert
^{\gamma_{JCF}}%
\end{equation}
\noindent where the gamma factor $\gamma_{JCF}$ is chosen such that $K_{JCF}=K_{DAS}$, and so forth for other beamforming methods. In this manner, images from different methods are compared under the constraint that their relative pixel variations are kept the same. The technique of contrast matching achieves a similar result as what is known as partial histogram matching \cite{bottenus_histogram_2021}, though in a different manner. In effect, for all displayed images we match the ratios of their histogram widths to their histogram means. While this method of contrast matching fails to provide a quantitative measure as does gCNR, it does facilitate the qualitative comparison of images by providing a level of standardization between them.

\subsection{Experiments}
Results presented in this paper were generated by data obtained either experimentally or from an online repository. Experimental data was obtained using a Verasonics Vantage 256 system. Data processing was performed in Matlab. We wrote a custom C++ beamformer using the open-source Eigen library \cite{eigenweb} inspired by the approach presented in the Matlab Ultrasound Toolbox (MUST) \cite{garcia_make_2021,perrot_so_2021}. We also implemented most of the algorithms described in Tables \ref{tableEQ} and \ref{tableEQ_part2} using C++ and Eigen, according to similar Matlab implementations from the Ultrasound Toolbox \cite{rodriguez-molares_ultrasound_2017}. All displayed images were compressed according to the contrast matching procedure described above, for standardized comparison.

\begin{figure*}[!t]
\centering
\includegraphics[width= \textwidth]{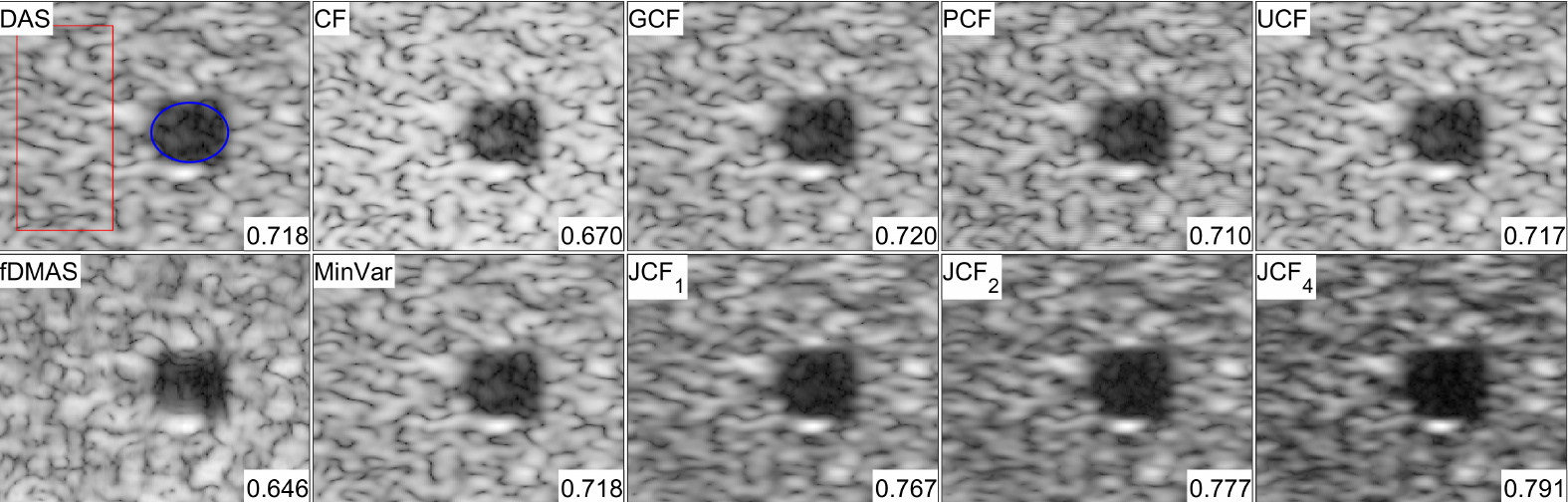}
\caption{Contrast-matched images of anechoic target in CIRS 040GSE phantom (2.0 mm diameter, 15 mm deep). JCF images are shown with increasing values of the smoothness parameter $\alpha$, leading to increasing suppression of noise and clutter. gCNR values associated with signal and background regions indicated respectively by blue circle and red rectangle in DAS panel are displayed at the bottom right of each panel.}
\label{figInclusion}
\end{figure*}

\begin{figure*}[!t]
\centering
\includegraphics[width= \textwidth]{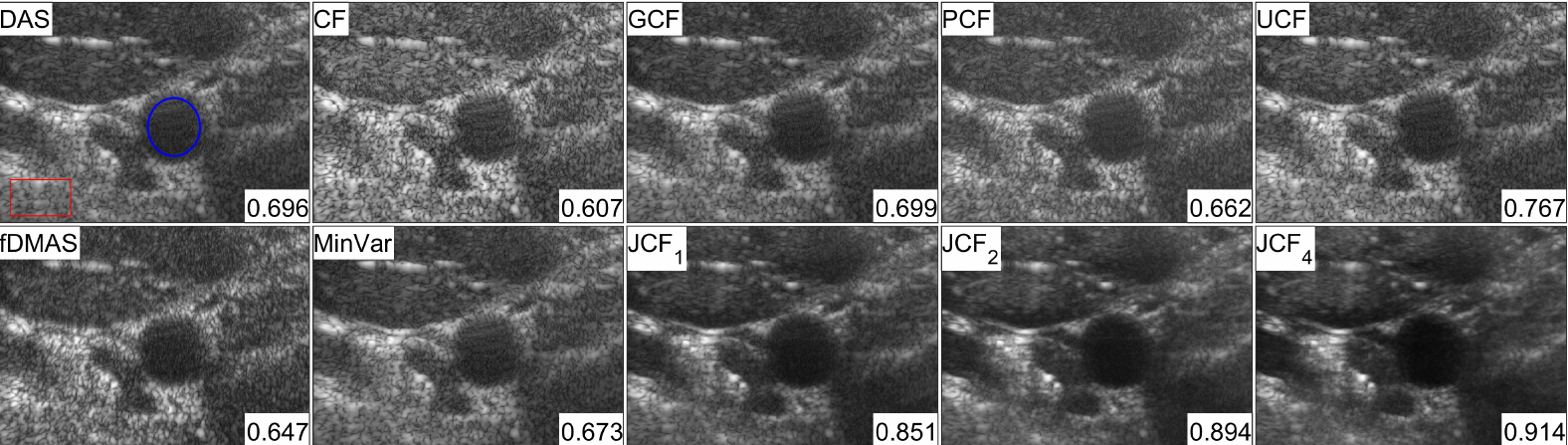}
\caption{Contrast-matched images beamformed from open-source raw data of a carotid artery (cross-sectional view) made available through the PICMUS challenge \cite{liebgott_plane-wave_2016}. JCF images are shown with increasing values of the smoothness parameter $\alpha$, leading to increasing suppression of noise and clutter. Image size is 32 mm x 22 mm (Horizontal x Vertical). gCNR values associated with signal and background regions indicated respectively by blue circle and red rectangle in DAS panel are displayed at the bottom right of each panel.}
\label{figCarotid}
\end{figure*}

\begin{figure*}[!t]
\centering
\includegraphics[width= \textwidth]{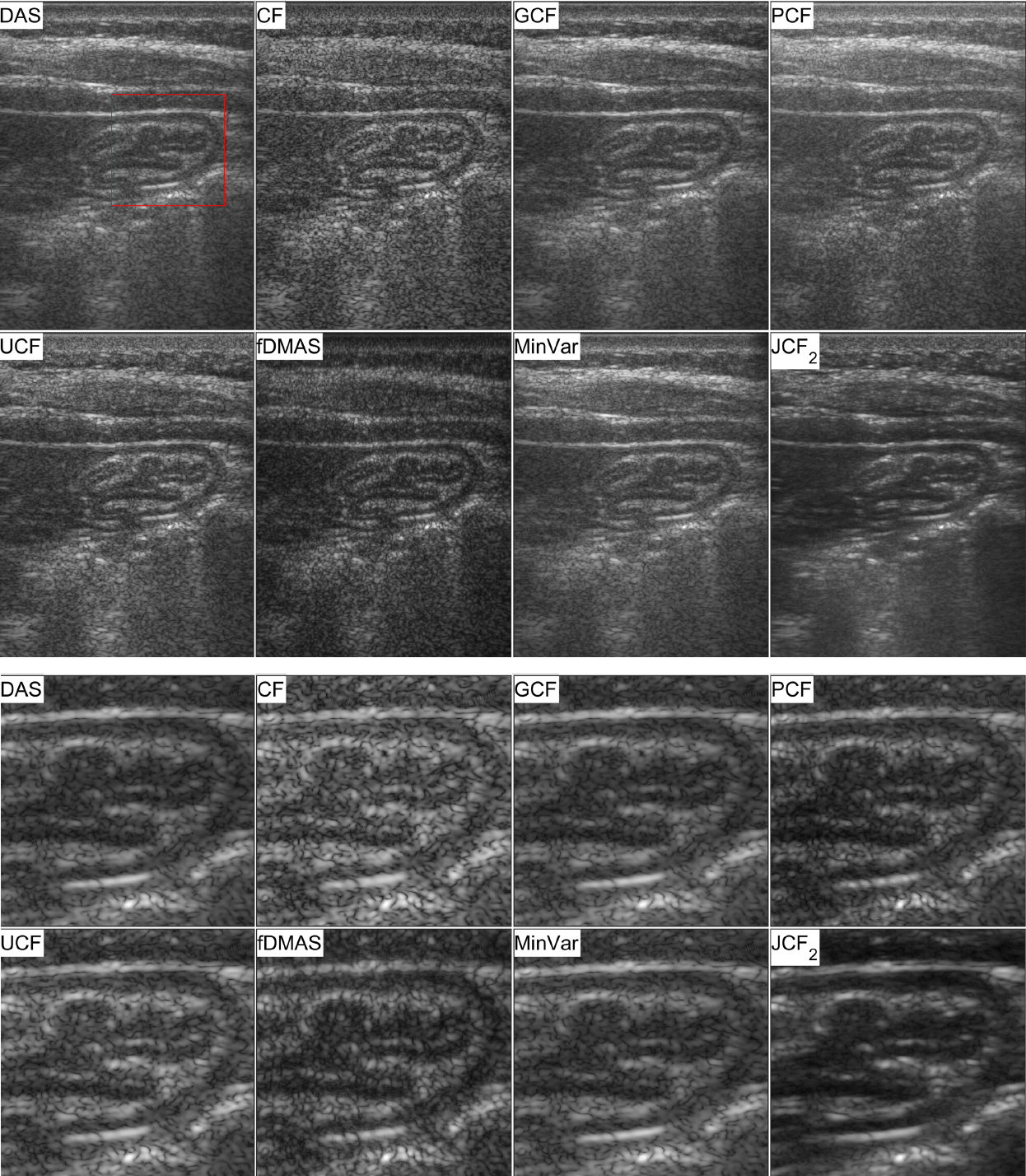}
\caption{Top: Contrast-matched images of stomach tissue from a healthy volunteer.  Image size is 44 mm x 57 mm (Horizontal x Vertical). Bottom: Contrast-matched images of magnified region indicated by red box in DAS panel above. Image size is 19 mm x 19 mm (Horizontal x Vertical).}
\label{fig_lung}
\end{figure*}

\begin{figure*}[!t]
\centering
\includegraphics[width= \textwidth]{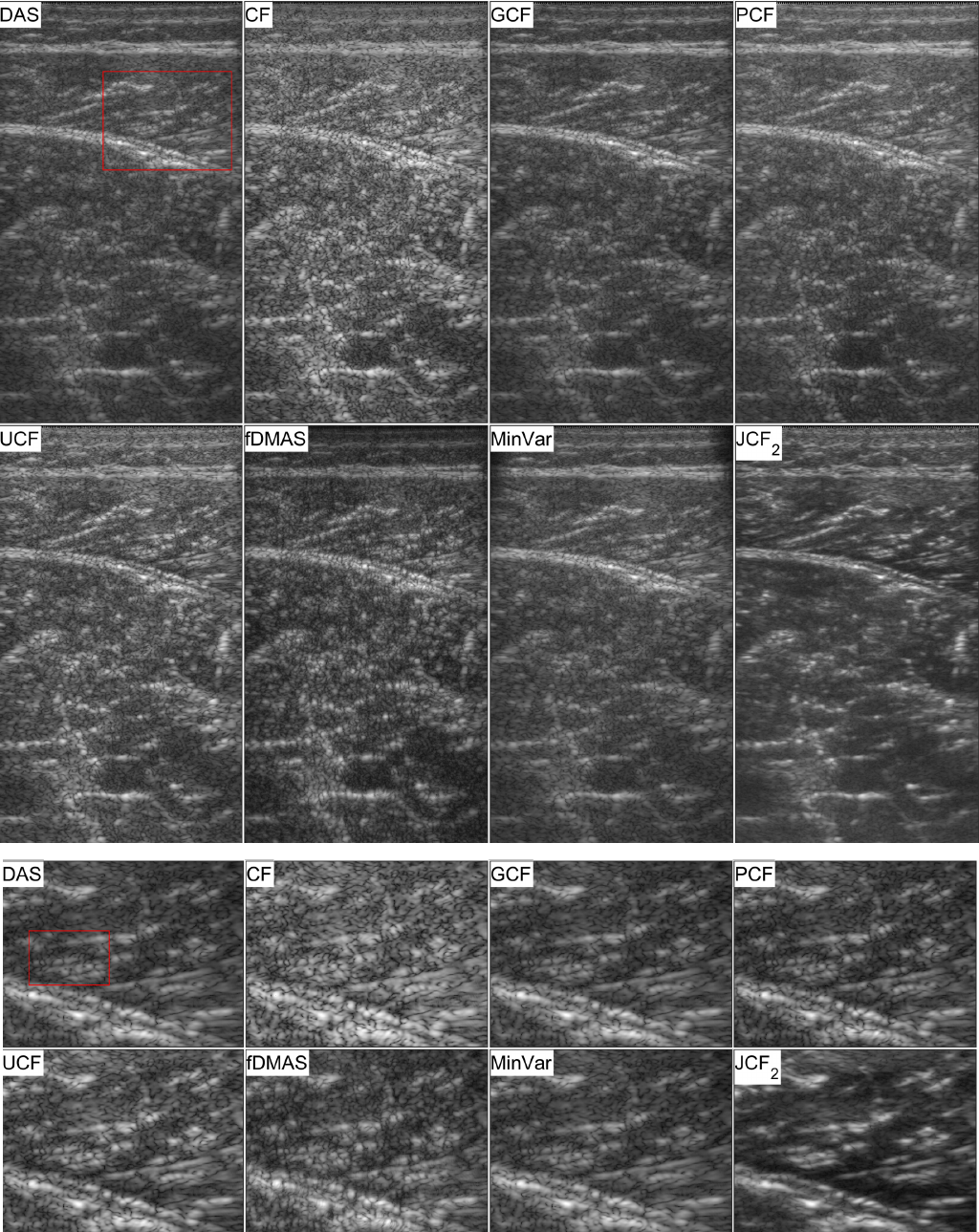}
\caption{Top: Contrast-matched images of calf muscle from a healthy volunteer.  Image size is 25 mm x 44 mm (Horizontal x Vertical). Bottom: Contrast-matched images of magnified region indicated by red box in DAS panel above.  Image size is 13 mm x 10 mm (Horizontal x Vertical).}
\label{fig_calf}
\end{figure*}

\begin{figure*}[!t]
\centering
\includegraphics[width= \textwidth]{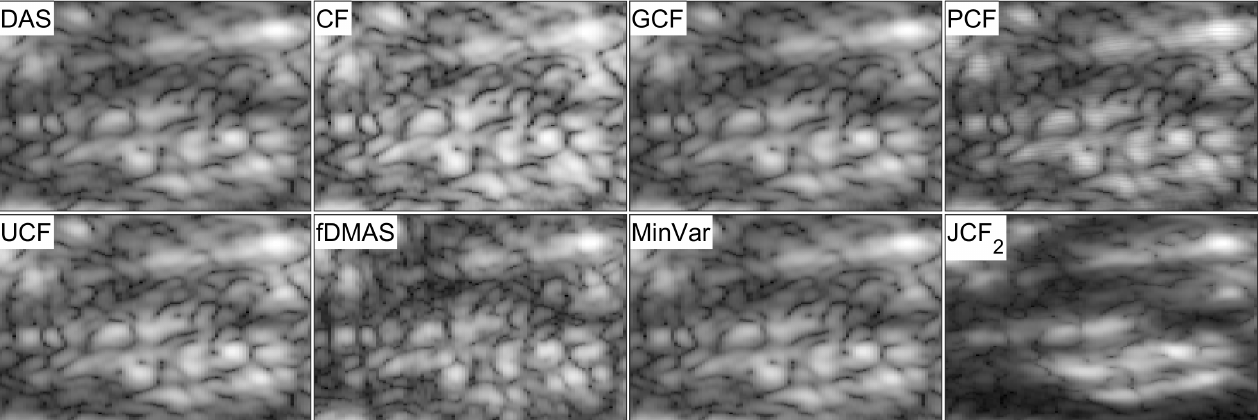}
\caption{Continuation of Fig. \ref{fig_calf} with increasing magnification of region indicated by red box in DAS panel. All panel sets are contrast-matched to respective DAS panel. JCF suppresses patchiness and distortions caused by speckle noise and clutter. Image size is 5 mm x 3 mm (Horizontal x Vertical).}
\label{fig_calf_zoom}
\end{figure*}

\begin{figure*}[!t]
\centering
\includegraphics[width= \textwidth]{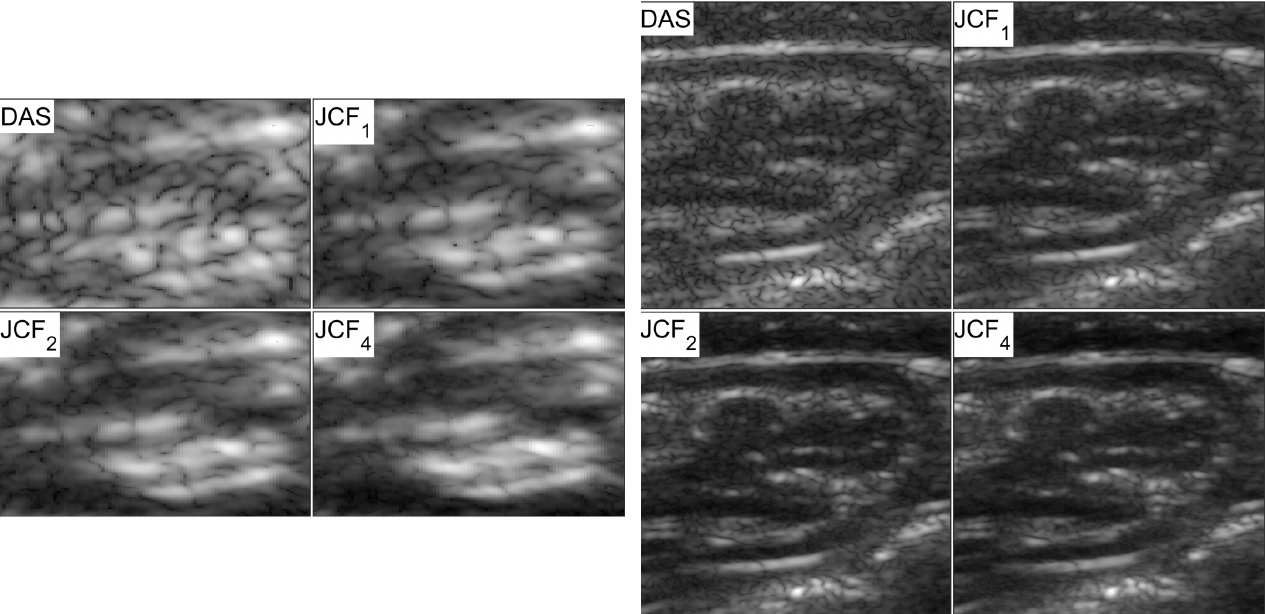}
\caption{Effect of change in smoothness parameter $\alpha$ on image quality ($\alpha=0$ is equivalent to DAS). Left: same view as Fig. \ref{fig_calf_zoom}. Right: same view as Fig. \ref{fig_lung}, bottom. Panels are contrast-matched. Increasing $\alpha$ increases noise and clutter suppression with little effect on resolution.}
\label{fig_variation}
\end{figure*}

\section{Results}
To begin, we imaged resolution point targets in a standard tissue-mimicking phantom (CIRS 040GSE Multi-Purpose Multi-Tissue Ultrasound Phantom) using a GE9LD probe (192 elements, element size 0.23 mm, center frequency 5.2 MHz). CPWC data was obtained using a sequence of 75 plane waves of angles $\theta$ uniformly distributed from -24 To 24 degrees. The results comparing several different beamforming modalities are shown in Fig. \ref{figRes}, where JCF$_2$ indicates $\alpha$ was set to 2. Plots of the displayed intensities across the indicated red lines are shown in the bottom panel. Several takeaways can be inferred from this plot. The first is that all the filtering techniques presented here produce more or less similar results compared to DAS in terms of target resolution (peak widths) and linearity (peak heights across the plots). However, when comparing the background noise throughout the images, we find that JCF$_2$ provides a consistently lower and smoother baseline than the other techniques. The reason for this is that the background here comes from speckle noise resulting from sub-resolution scatterers. Because scatterers are randomly distributed within each resolution volume, each transmit/receive combination produces a randomly varying signal element $s_{m,n}$ bearing little coherence relative to neighboring signal elements. When summed coherently, the signal elements produce highly noisy speckle of amplitude distribution generally defined by Rayleigh statistics \cite{goodman_speckle_2007}. JCF effectively attenuates such weakly coherent noise, manifestly more so than other techniques. 

As noted above, the gCNR metric is generally of limited practical utility. However, in certain cases it is possible to unambiguously draw a line between signal and background. An example of this is in a phantom sample (e.g. CIRS 040GSE), where the presence of inclusions is well defined. Such an example is shown in Fig. \ref{figInclusion} where an anechoic inclusion (blue circle) is clearly visible within its surrounding background (red rectangle). Another example comes from imaging a carotid artery. Here, too, the anechoic artery lumen can be easily identified. Figure \ref{figCarotid} shows images obtained from an open-source CPWC dataset of a carotid artery from the PICMUS challenge \cite{liebgott_plane-wave_2016}. For both the phantom and carotid artery images we illustrate the effect of different JCF variants where $\alpha$ ranges from $1$ to $4$. We emphasize that the overall contrast of each panel, as defined by Eq. \ref{contrast}, is rigorously kept the same across panels, and yet the contrast of the JCF panels \textit{appears} to be higher than the rest (the apparent contrast increasing with $\alpha$). The reason for this is that JCF filtering suppresses noise and clutter better than the other modalities. As such, the pixels that contribute to contrast in the JCF images are dominantly signal pixels, whereas much of the contrast from the other modalities originates from noise or clutter. An examination of the artery lumen further reveals that the streak artifacts apparent with other modalities all but disappear in the JCF panels, particularly as $\alpha$ increases. These qualitative observations are borne out by the quantitative gCNR measures distinguishing signal from background. In agreement with \cite{rodriguez-molares_generalized_2020}, we find that alternative coherence-based modalities do not significantly enhance gCNR. In contrast, JCF filtering does manifestly lead to a significant increase in gCNR. This increase in gCNR originates entirely from JCF filtering, since contrast matching itself does not affect gCNR.  

As will be shown below, JCF is particularly effective at improving the quality of soft-tissue imaging. Figure \ref{fig_lung} shows images of stomach tissue obtained from a healthy volunteer. Again we used a GE9LD probe, this time with 45 plane-wave transmits of angles uniformly distributed between -10 and 10 degrees. Again, the noise and clutter in the JCF panel appear smoother and more suppressed compared to the rest. This is more apparent when the images are magnified (zoomed region delimited by the red box in the DAS panel). 

Figures \ref{fig_calf} and \ref{fig_calf_zoom} show images of calf muscle tissue from the same volunteer, acquired this time with a L12-3v probe (128 elements, element size 0.2 mm, center frequency ~9 MHz) using 45 plane-wave transmits of angles uniformly distributed between -18 and 18. Two zoom levels are shown of increasing magnification (red boxes). The benefits of JCF are most apparent at higher magnification (Fig. \ref{fig_calf_zoom}). Whereas the images from all other modalities are marred by considerable speckle noise that masks the underlying sample structure, the JCF panel is largely devoid of this noise. As a result, the sample structure (here perimysial septa \cite{albayda_diagnostic_2020}) appears less distorted and mottled. These panels in particular highlight the deficiency in using contrast alone as a measure of image quality. The contrasts of all panels in Fig. \ref{fig_calf_zoom} are matched to be the same, and yet the image quality provided by JCF appears manifestly better. 

Finally, Fig. \ref{fig_variation} shows side-by-side comparisons of the effect of increasing the smoothness parameter $\alpha$, compared to conventional DAS (recalling that DAS is equivalent to JCF$_0$). The transition from $\alpha=0$ to $\alpha=1$ provides the most apparent benefit in image quality. This trend continues with increasing $\alpha$, though with diminishing returns in imaging quality and noise suppression (as confirmed quantitatively by the diminishing increase in gCNR in Figs. \ref{figInclusion} and \ref{figCarotid}). Notably, increasing $\alpha$ does not appear to undermine spatial resolution in any way, nor does it appear to alter the relative brightnesses of the signals themselves.

\section{Discussion}
The basic principle of JCF is to individually weight every signal element $s_{m,n}$ within the full transmit/receive matrix $\mathbf{S}$ according to its respective signal quality. The weighting $w_{m,n}$ is adaptive in the sense that the signal qualities are derived from the measurements themselves rather than from a predefined model. That is, different combinations of transmit angle (indexed by $m$) and receive position (indexed by $n$) can produce different degrees of signal quality based more or less on happenstance, depending on a variety of factors that, in general, cannot be predicted or controlled in advance. Signal quality for each transmit/receive combination $(m,n)$ is determined by the degree of coherence measured jointly across both the associated transmit angles (column $n$) and associated receive positions (row $m$). The jointness of this measure is a defining feature of JCF-assisted CPWC beamforming. 

Different factors can affect the measured coherence of each signal element. For example, if measurements are dominated by clutter or electronic noise, then the elements $s_{m,n}$ are dominantly uncorrelated with one another and $w_{m,n}$ becomes correspondingly low. Similarly, if random sample micro-motion occurs between each transmit event, this too leads to row and/or column decorrelation between $s_{m,n}$ elements, also leading to low $w_{m,n}$. Another source of decorrelation can come from upstream sample occlusions or aberrations that can cause changes in amplitude or phase that are different for different $(m,n)$ combinations. Yet another source of decorrelation comes from the presence of randomly distributed sub-resolution scatterers, as discussed related to Fig. \ref{figRes}, leading to speckle noise. Regardless of the sources of decorrelations, their resulting effects on signal quality are considered deleterious. JCF makes no attempt to correct for these effects, but rather suppresses them according to their severity. As a result, the remaining signal elements that contribute to the coherent summation in CPWC feature high degrees of row/column coherence, leading to noticeable improvements in image quality.

The benefits in image quality that we report here for soft-tissue imaging are more qualitative than quantitative. The reason for this stems from the difficulty in defining a quantitative metric for image quality that does not rely on a priori knowledge of the tissue. Nevertheless, we established a methodology based on contrast matching to provide a reasonably level field for comparison between different CPWC beamforming modalities. This methodology makes use of gamma compression with gamma values adjusted to yield a consistent predefined image contrast (ratio of histogram width to mean) across all modalities. In our case, we chose the DAS image contrast to serve as the predefined reference contrast. For all our results, we purposefully chose a somewhat low $\gamma_{DAS}$ value of 0.25 to better reveal the weak intensities in our displayed images, where the benefits of JCF are most apparent. A different reference gamma value could, of course, also be used.

Regarding the benefits of JCF, the most significant is the suppression of uncorrelated noise or clutter in the beamformed images, leading to an increase in \textit{apparent} image contrast that better enables the distinction between signal and background, at least qualitatively. JCF uses the same raw data as conventional DAS CPWC, whith no requirement of additional measurements or a priori knowledge about the sample. However, the benefits do not come without trade-offs. There is an additional computational load associated with JCF. In our case, there was also an additional requirement of computer memory. Our specific implementation was tailored for post-processing of saved, time-delayed data rather than for real-time processing. This approach involved saving the entire 4D $\mathbf{S}(x,z)$ matrix instead of calculating a 2D $\mathbf{S}$ matrix for each $(x,z)$ sample coordinate on the fly. With our system (Intel Xeon E5-2680 2.5 GHz CPU, 12 cores, 24 threads, 128 GB DDR4 RAM), JCF beamforming was approximately 3-4 times slower than standard DAS beamforming. We anticipate that a more optimized, pixel-wise approach and/or GPU-based implementation could significantly speed up JCF beamforming and enable live imaging applications. A Matlab version of our code is available at
https://github.com/biomicroscopy/Ultrasound JCF.

\section{Conclusion}
We have described a method to identify and suppress uncorrelated noise in CPWC beamforming. The method involves a fine-grained adaptive filtering at the level of the individual signal elements contributing to conventional DAS coherent summation, based on their local spatio-angular joint coherence. The benefits are a reduction and smoothing of background noise and clutter, leading to improved image quality and, when applicable, improved gCNR. The benefits are apparent, for example, when images are corrupted by random patchiness caused by speckle noise, which can severely distort and mask the underlying signals of interest. JCF effectively attenuates this noise without noticeable loss in spatial resolution. We further introduced a general contrast-matching approach, similar to partial histogram matching, to facilitate the comparison of JCF with other methods of image filtering. The aim of this work is to add JCF to the arsenal of CPWC beamforming strategies, with the hope of improving ultrasound imaging quality for medical applications.  

\section*{Acknowledgments}
All images of human volunteers were acquired in accordance with IRB Protocol \# 5914E approved by the Boston University Institutional Regulatory Board. We thank Thomas Bifano and the BU Photonics Center for making a Verasonics machine available to us.

\bibliographystyle{IEEEtran}
\bibliography{JCFrefs,EigenRef}

\begin{table*}[htbp]
    \caption{Method and references (Part I)}
    \label{tableEQ}
    \setlength{\tabcolsep}{3pt}
    \begin{tabular}{p{100pt} p{300pt} p{100pt}}
    \hline\hline
    Method Name & 
    Expression & References \\[1.0em]
    \hline
    DAS &
        \begin{minipage}[c]{.3\textwidth}
        \[
             B_{\text{das}}(\vec{r}) = \frac{1}{MN} \left( \sum_{m=1}^{M} \sum_{n=1}^{N} s_{mn}(\vec{r}) \right) 
        \]
        \end{minipage} & \cite{montaldo_coherent_2009,perrot_so_2021,rindal_effect_2019,rodriguez-molares_generalized_2020} \\[4.0em] \hline
    CF & 
        \begin{minipage}[c]{.3\textwidth}
        \[
            \renewcommand{\arraystretch}{2.5} % Line spacing val
            \large{ % Font size command
            \begin{array}{ll}
                w_m(\vec{r}) &= \frac{\left|\sum_{n=1}^{N}s_{mn}\right|^2}{N\sum_{n=1}^{N}|s_{mn}|^2}, \\
                B_{\text{CF}}(\vec{r}) &= \frac{\sum_{m=1}^M\sum_{n=1}^{N}{w_ms_{mn}}}{N}
            \end{array}
            }
        \]
        \end{minipage} & \cite{mallart_van_1991,mallart_adaptive_1994,hollman_coherence_1999,rindal_effect_2019,rodriguez-molares_generalized_2020} \\[4.0em] \hline
    GCF & 
        \begin{minipage}[c]{.3\textwidth}
        \[
            \renewcommand{\arraystretch}{2.5} % Line spacing val
            \large{ % Font size command
            \begin{array}{ll}
                w_m(\vec{r}) &= \frac{\sum_{n\in L\ F}\ s_{mn}}{\sum_{n}|s_{mn}|^2}, \\
                B_{\text{GCF}}(\vec{r}) &= \frac{\sum_{m=1}^M\sum_{n=1}^{N}{w_ms_{mn}}}{N}
            \end{array}
            }
        \]
        \end{minipage} & \cite{li_adaptive_2003,rindal_effect_2019,rodriguez-molares_generalized_2020} \\[4.0em] \hline
    PCF & 
        \begin{minipage}[c]{.3\textwidth}
        \begin{align*}
            \renewcommand{\arraystretch}{2.5} % Line spacing val
            \large{ % Font size command
            \begin{array}{ll}
                p_m &= \min\left\{\sigma(\arg(s_{mn}))_n,\sigma(\arg\text{unwrap}(s_{mn}))_n\right\}, \\
                w_m(\vec{r}) &= \max\left\{0,1-\frac{\gamma}{\sigma_0}p_m\right\}, \\
                B_{\text{PCF}}(\vec{r}) &= \frac{\sum_{m=1}^M\sum_{n=1}^{N}{w_ms_{mn}}}{N}
            \end{array}
            }
        \end{align*}
        \end{minipage} & \cite{camacho_phase_2009,rindal_effect_2019,rodriguez-molares_generalized_2020} \\[4.0em] \hline
    UCF & 
        \begin{minipage}[c]{.3\textwidth}
        \[
            \renewcommand{\arraystretch}{2.5} % Line spacing val
            \large{ % Font size command
            \begin{array}{ll}
                w(\vec{r}) &= \frac{\left|\sum_{m=1}^{M}\sum_{n=1}^{N}s_{mn}\right|^2}{MN\sum_{m=1}^{M}\sum_{n=1}^{N}|s_{mn}|^2}, \\
                B_{\text{UCF}} &= w\sum_{m=1}^{M}\sum_{n=1}^{N}s_{mn}
            \end{array}
            }
        \]
        \end{minipage} & \cite{yang_coherent_2020,yang_united_2020} \\[4.0em] \hline
    fDMAS & 
    \begin{minipage}[c]{.3\textwidth}
        \[
            \renewcommand{\arraystretch}{2.5} % Line spacing val
            \large{ % Font size command
             \begin{array}{ll}
                s_n\left(\vec{r}\right)=\sum_{m=1}^{M}\mathbb{R}\left\{s_{mn}\left(\vec{r}\right)\right\}, \\
                {\hat{s}}_n\left(\vec{r}\right)=sign\left(s_n\left(\vec{r}\right)\right)\sqrt{s_n\left(\vec{r}\right)}, \\

                B_{DMAS}\left(\vec{r}\right)=\sum_{l=1}^{L}\sum_{n=1}^{N-l}{{\hat{s}}_n\left(\vec{r}\right){\hat{s}}_{n+l}\left(\vec{r}\right)}
            \end{array}
            }
        \]
        \end{minipage} & \cite{matrone_delay_2015,rindal_effect_2019,rodriguez-molares_generalized_2020} \\[4.0em]
    \hline\hline
    \multicolumn{2}{p{400pt}}{Algorithms used for the methods compared in this paper (continued in Table \ref{tableEQ_part2}).}
    \end{tabular}
\end{table*}

\begin{table*}[htbp]
    \caption{Methods and references (Part II)}
    \label{tableEQ_part2}
    \centering
    \setlength{\tabcolsep}{3pt}
    \begin{tabular}{p{100pt} p{300pt} p{100pt}}
    \hline\hline
    Method Name & Expression & References \\
    \hline
    MinVar & 
    \begin{minipage}[c]{.3\textwidth}
        \[
            \renewcommand{\arraystretch}{2.5} % Line spacing val
            \large{ % Font size command
             \begin{array}{ll}
                s_n\left(\vec{r}\right)=\sum_{m=1}^{M}{s_{mn}\left(\vec{r}\right)}, \\
                {\bar{s}}_n\left(\vec{r}\right)=\left[s_n\left(\vec{r}\right)s_{n+1}\left(\vec{r}\right)\cdots\cdots s_{n+L-1}\left(\vec{r}\right)\right]^T,\\

                \hat{R}\left(\vec{r}\right)=\frac{\sum_{k=-K}^{K}\sum_{n=0}^{N-L}{{\bar{s}}_n\left(\vec{r}=\left(x,z-k\right)\right){\bar{s}}_n^H\left(\vec{r}=\left(x,z-k\right)\right)}}{\left(2K+1\right)\left(N-L+1\right)},\\

                \widetilde{R}\left(\vec{r}\right)=\hat{R}\left(\vec{r}\right)+\frac{\Delta}{L}tr{\left\{\hat{R}\left(\vec{r}\right)\right\}I},\\

                w_{MV}\left(\vec{r}\right)=\frac{{\widetilde{R}}^{-1}\vec{a}}{{\vec{a}}^H{\widetilde{R}}^{-1}a},\\

                B_{MV}\left(\vec{r}\right)=\frac{1}{N-L+1}\sum_{l=0}^{N-L}{w_{MV}^H\left(\vec{r}\right){\bar{s}}_n\left(\vec{r}\right)}

            \end{array}
            }
        \]
        \end{minipage} & \cite{synnevag_benefits_2009,asl_eigenspace-based_2010,lan_joint_2021,rindal_effect_2019,rodriguez-molares_generalized_2020} \\[4.0em] \hline
    JCF & 
    \begin{minipage}[c]{.3\textwidth}
        \[
            \renewcommand{\arraystretch}{2.5} % Line spacing val
            \large{ % Font size command
             \begin{array}{ll}
                w_{m,n}\left(  x,z\right)  =\frac{\left\vert \sum\limits_{m^{\prime}=1}%
                ^{M}\sum\limits_{n^{\prime}=1}^{N}s_{m^{\prime},n}s_{m,n^{\prime}}\right\vert
                ^{\alpha}}{\left(  MN\right)  ^{\alpha-1}\sum\limits_{m^{\prime}=1}^{M}%
                \sum\limits_{n^{\prime}=1}^{N}\left\vert s_{m^{\prime},n}\right\vert ^{\alpha
                }\left\vert s_{m,n^{\prime}}\right\vert ^{\alpha}} \\
                B_{JCF}\left(  x,z\right)  =\frac{1}{MN}\sum\limits_{m=1}^{M}\sum\limits_{n=1}^{N}w_{m,n}s_{m,n}
            \end{array} 
            }
        \]
        \end{minipage} \\[4.0em]
    \hline\hline
    \multicolumn{2}{p{400pt}}{Continuation of Table \ref{tableEQ}.}
    \end{tabular}
\end{table*}

\end{document}